# Geoglyphs of Titicaca as an ancient example of graphic design

**Amelia Carolina Sparavigna**
Dipartimento di Fisica, Politecnico di Torino
C.so Duca degli Abruzzi 24, Torino, Italy

**Abstract**
The paper proposes an ancient landscape design as an example of graphic design for an age and place where no written documents existed. It is created by a network of earthworks, which constitute the remains of an extensive ancient agricultural system. It can be seen by means of the Google satellite imagery on the Peruvian region near the Titicaca Lake, as a texture superimposed to the background landform. In this texture, many drawings (geoglyphs) can be observed.

**Keywords**: Geoglyphs, History of Graphics, Image processing, Satellite maps, Archaeology

According to the Encyclopaedia Britannica, graphic design is the art and profession of selecting and arranging visual elements. It is a creative process used to convey a message to a targeted audience, involving many artistic and professional disciplines, focussed on visual communication. In the same article of this Encyclopaedia, we read that "the evolution of graphic design as a practice and profession has been closely bound to technological innovations, societal needs, and the visual imagination of practitioners". For this reason, it is natural for us to assume graphic design applied to logos and branding, web sites, magazines, newspapers, and packaging.
The article continues telling that "graphics has been practiced in various forms throughout the history", because strong examples can be found in ancient manuscripts [1], in Egypt and China. In fact, before what we are calling "history", there were ages when humankind dealt with stone, wood or bone tools only. We can ask ourselves what could be a "graphic design" for those ancient populations and which possible substrates used for it.
The archaeological findings are providing some answers which are coming only from those objects conserved on non-perishable substrates. Let us choose a well-known case, the Lascaux Cave paintings: in spite of a huge elapse of time, when observing its drawn animals and symbols, we see them quite modern and we are naturally driven to consider them as an effective example of graphic communication. There are many proposed meanings for Lascaux drawings: prehistoric star charts, visions experienced during ritualistic trances, accounts of past hunting success, mystical rituals for improving hunt, illustrated myths [2]. In all the cases, the drawings are messages between human world and spirit worlds.
Once agricultural civilizations flourished, in some places human kind passed from the use of symbols to a writing system. There, the history of graphics began, because we have the early objects incorporating signs, symbols, icons and written texts (rather interesting examples of graphics are the icons on Harappa seals, [3]).
In other cases, agricultural civilizations flourished but a writing system was not developed. Moreover, no statues, no painting, few buildings remain. Nevertheless, there is a substrate that humankind used for creating designs and this substrate is the landscape. To modify the surface of the ground, in South America, ancient populations used the earthworks. These manmade structures are appearing as a texture superimposed to the background landform. In some cases, they remain quite visible, widened and flattened as a consequence of the natural degradation processes (unfortunately, these structures cannot survive the action of men).

In this case, it is the technology answering to our question about graphics of those populations, with the analysis of satellite imagery. A wonderful artificial landform, that can be observed with Google Maps imagery is that created by a huge network of earthworks covering land and hills near the Titicaca Lake, being the result of the agricultural effort of the ancient Andean people [4,5]. People created a system of raised fields, which were large elevated planting platforms, with the corresponding drainage canals. This system improved soil conditions, the temperature and moisture conditions for crops. Besides the earthworks of the plain areas, we can see hill slopes criss-crossed with terrace walls.

Lake Titicaca sits in a basin high in the Andes on the border of Peru and Bolivia. The western part of the lake lies within the Puno Region of Peru, and the eastern side is located in the Bolivian La Paz Department. Both regions have terraced hill slopes. Some of the landforms near Lake Titicaca are rather remarkable, because they have a clear symbolic meaning [6]. The observation of geoglyphs in the Bolivian region of Titicaca was previously reported in [7,8]. To the author's knowledge, the geoglyphs of the Peruvian region are not shown in literature: it seems that Ref.6 is the first paper on the subject.

As discussed in Ref.6, many geoglyphs are representing animals (birds, snakes, a hedgehog, but there are also geometric drawings). After a survey of the region, we proposed a rule of thumb: to find the figures, look for circular ponds, because sometimes they can be the eye of an animal. This undoubtedly means that the drainage system of terraced hills and raised fields was carefully planned, with symbolic landforms at specific points of the system. As a consequence, this system was created from an engineering/landscape project created with a graphic design. It is also possible that the use of geoglyphs could possess an ideal function to protect the fields from floods, attracting the benevolent natural spirits.

Let us see an example of Peruvian geoglyphs. It is shown in Figure 1, from an original Google Maps image subjected to the processing method in Ref.9. The figure is representing a snake: to avoid any doubt, Figure 2 is showing the head in detail, where the bifid tongue is clearly displayed. It is author's opinion, that this snake is one element of a more complex drawing, with another animal, biting the snake (see Figure 3, the shape of the two animals in fight are highlighted in white).

In conclusion, let us remark two facts. First of all, that image processing of visible satellite imagery, and in particular of the Google Maps images, can represent a quite useful tool for archaeological and historical researches. Second, that the notion of graphics design can be extended to large scale objects, also in the case of very ancient structures.

Geophysics, arxiv:1009.2231, http://arxiv.org/abs/1009.2231
[7] Discovery of vast prehistoric works built by Giants? The Geoglyphs of Teohuanaco, David E. Flynn, a post of February 24, 2008
[8] http://www.atlantisbolivia.org/geoforms.htm
[9] Enhancing the Google imagery using a wavelet filter, A.C. Sparavigna, 8 Sept 2010. Geophysics; Earth and Planetary Astrophysics, arXiv:1009.1590v1

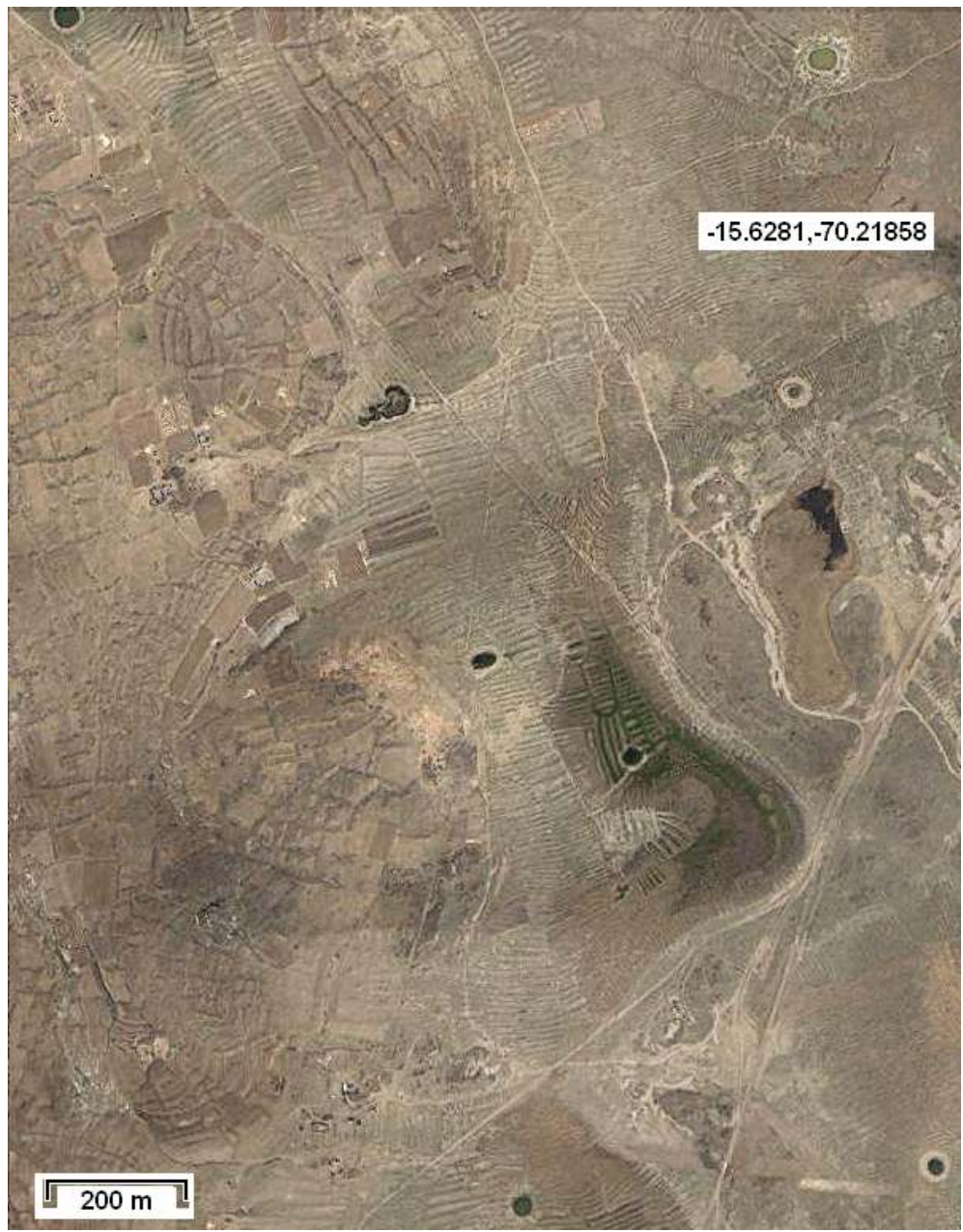

Fig.1 This is one among the Peruvian geoglyphs of Titicaca. It is chosen as an example of graphic design of the landform. It shows a snake, with the body and texture of skin created by the terraced hill (left part of the image), the head (darker area) on the plain surface of the ground, a pond as its eye. The head is shown in detail in the following image.

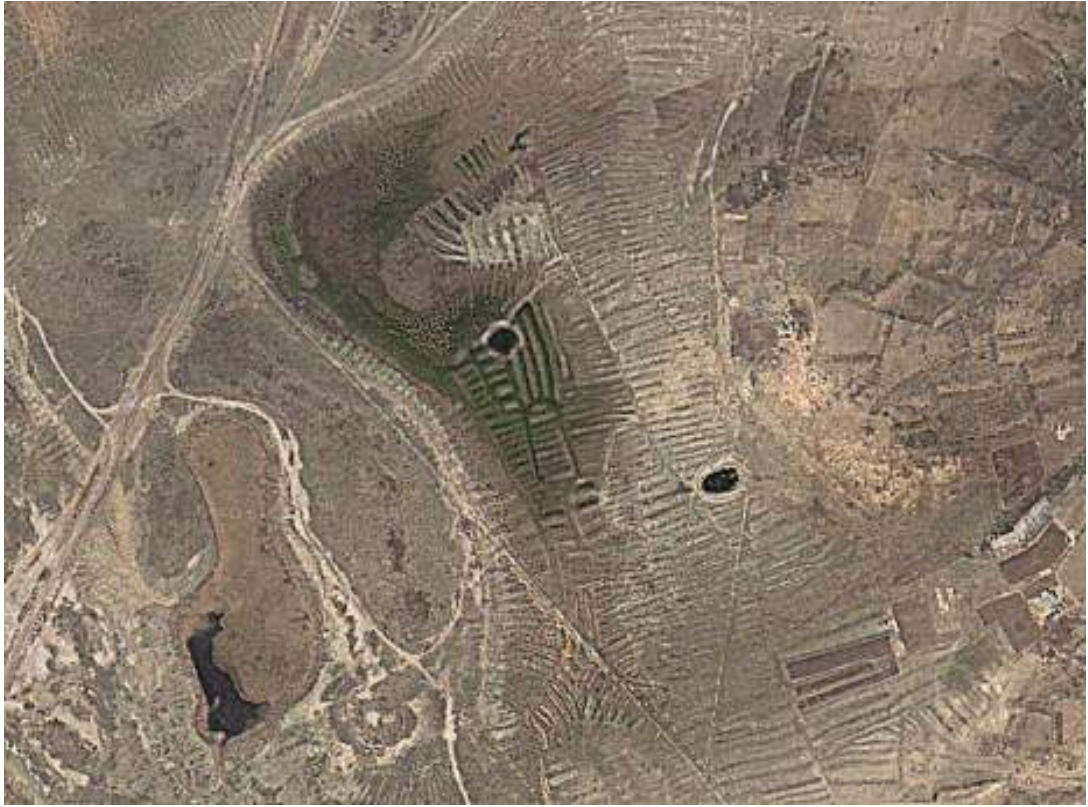

Fig.2 The detail of the head. To avoid any doubt, note the bifid tongue of the snake.

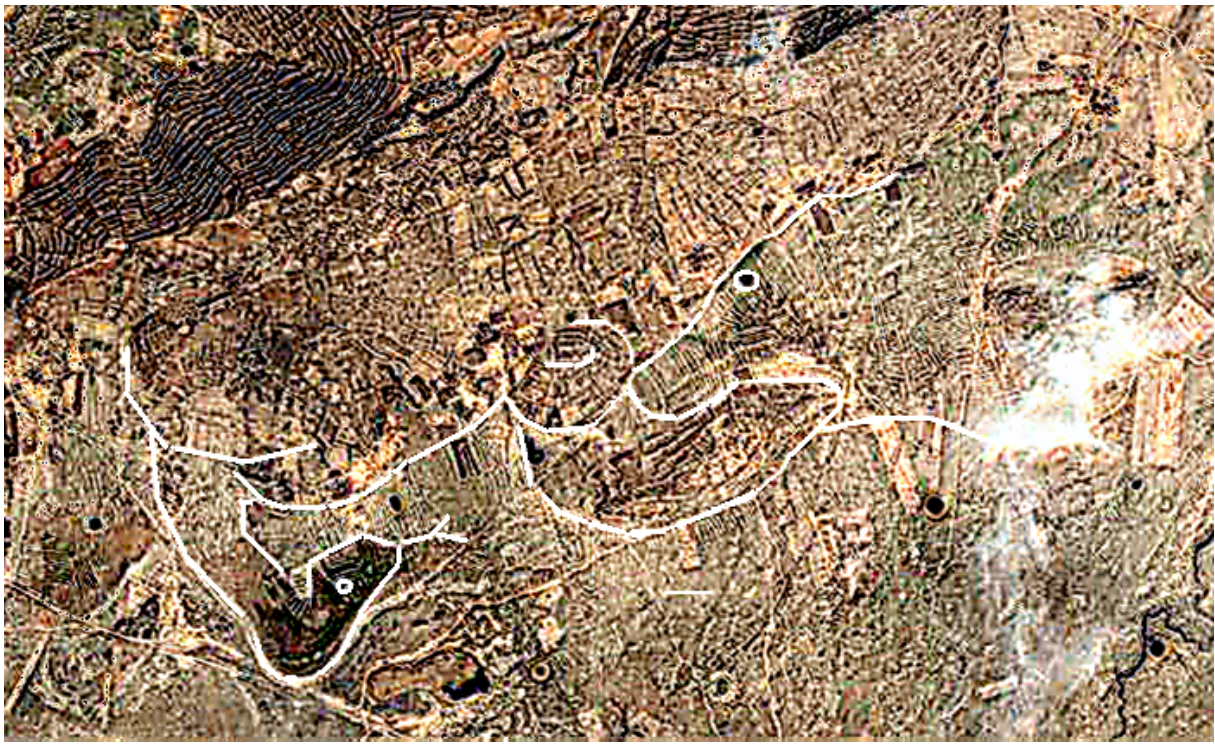

Fig.3 An animal is biting the body of the snake.